\documentclass[prl,showpacs,superscriptaddress,twocolumn]{revtex4}

\usepackage{amsmath,bm}
\usepackage{graphicx}
\usepackage{epstopdf}

\usepackage{color}
\begin{document}

\newcommand{\vk}{{\vec k}}
\newcommand{\vK}{{\vec K}}
\newcommand{\vb}{{\vec b}}
\newcommand{{\vp}}{{\vec p}}
\newcommand{{\vq}}{{\vec q}}
\newcommand{\vQ}{{\vec Q}}
\newcommand{\vx}{{\vec x}}
\newcommand{\beq}{\begin{equation}}
\newcommand{\eeq}{\end{equation}}
\newcommand{\half}{{\textstyle \frac{1}{2}}}
\newcommand{\gton}{\stackrel{>}{\sim}}
\newcommand{\lton}{\mathrel{\lower.9ex \hbox{$\stackrel{\displaystyle<}{\sim}$}}}
\newcommand{\ee}{\end{equation}}
\newcommand{\ben}{\begin{enumerate}}
\newcommand{\een}{\end{enumerate}}
\newcommand{\bit}{\begin{itemize}}
\newcommand{\eit}{\end{itemize}}
\newcommand{\bc}{\begin{center}}
\newcommand{\ec}{\end{center}}
\newcommand{\bea}{\begin{eqnarray}}
\newcommand{\eea}{\end{eqnarray}}

\newcommand{\beqar}{\begin{eqnarray}}
\newcommand{\eeqar}[1]{\label{#1} \end{eqnarray}}
\newcommand{\pleft}{\stackrel{\leftarrow}{\partial}}
\newcommand{\pright}{\stackrel{\rightarrow}{\partial}}

\newcommand{\eq}[1]{Eq.~(\ref{#1})}
\newcommand{\fig}[1]{Fig.~\ref{#1}}
\newcommand{\eff}{ef\!f}
\newcommand{\alphas}{\alpha_s}

\renewcommand{\topfraction}{0.85}
\renewcommand{\textfraction}{0.1}
\renewcommand{\floatpagefraction}{0.75}

\title{ $\eta$ meson production of high-energy nuclear collisions at NLO }

\date{\today  \hspace{1ex}}
\author{Wei Dai\footnote{daiw@mail.tsinghua.edu.cn}}
 \affiliation{Physics Department, Tsinghua University, Beijing, China}
\author{Xiao-Fang Chen}
\affiliation{School of Physics and Electronic Engineering, Jiangsu Normal University, Xuzhou 221116, China}

\author{Ben-Wei Zhang\footnote{bwzhang@mail.ccnu.edu.cn}}
\affiliation{Key Laboratory of Quark \& Lepton Physics (MOE) and Institute of Particle Physics,
 Central China Normal University, Wuhan 430079, China}
\author{Enke Wang}

\affiliation{Key Laboratory of Quark \& Lepton Physics (MOE) and Institute of Particle Physics,
 Central China Normal University, Wuhan 430079, China}
\begin{abstract}

The transverse momentum spectrum of $\eta$ meson in relativistic heavy-ion collisions is studied at the Next-to-Leading Order (NLO) within the perturbative QCD, where the jet quenching effect in the QGP is incorporated with the effectively medium-modified $\eta$ fragmentation functions using the higher-twist approach. We show that the theoretical simulations could give nice descriptions of PHENIX data on $\eta$ meson in both $\rm p+p$ and central $\rm Au+Au$ collisions at the RHIC, and also provide numerical predictions of $\eta$ spectra in central $\rm Pb+Pb$ collisions with $\sqrt{s_{NN}}=2.76$~TeV at the LHC.  The ratios of $\eta/\pi^0$ in $\rm p+p$ and in central $\rm Au+Au$ collisions at $200$~GeV are found to overlap in a wide $p_T$ region, which matches well the measured ratio $\eta / \pi^0$ by PHENIX. We demonstrate that, at the asymptotic region when $p_{T} \rightarrow \infty$ the ratios of $\eta/\pi^{0}$ in both $\rm Au+Au$ and $\rm p+p$ are almost determined only by quark jets fragmentation and thus approach to the one in $e^{+} e^{-}$ scattering; in addition, the almost identical gluon (quark) contribution fractions to $\eta$ and to $\pi$ result in a rather moderate variation of $\eta/\pi^{0}$ distribution at intermediate and high $p_T$ region in $\rm A+A$ relative to that in $\rm p+p$; while a slightly higher $\eta/\pi^{0}$ at small $p_T$ in $\rm Au+Au$ can be observed due to larger suppression of gluon contribution fraction to $\pi^{0}$ as compared to the one to $\eta$. The theoretical prediction for $\eta / \pi^0$ at the LHC has also been presented.

\end{abstract}

\

\maketitle

The strong suppression of single hadron production at large transverse momentum~\cite{star-suppression,phenix-suppression} has provided the convincing  evidence of the jet quenching phenomena discovered in relativistic heavy-ion collisions (HIC)~\cite{Gyulassy:2003mc}. Extensive phenomenological investigations~\cite{Vitev:2002pf,Wang:2003mm,Eskola:2004cr,Renk:2006nd,Chen:2010te,Chen:2011vt} and experimental measurements~\cite{Adare:2012wg, Adare:2012uk, Afanasiev:2009aa, Adler:2006bw, Adler:2003qi,Abelev:2009wx,Abelev:2007ra} on the suppression of single hadron spectra at high $p_T$ have been carried out at both the RHIC and the LHC.  As the first observable of jet quenching phenomena, the yield suppression of inclusive hadrons is arguably the most thoroughly studied quantity of jet quenching,  and provides an indispensible tool to extract the properties of the hot medium created in nucleus-nucleus collisions by comparing theoretical calculations with experimental measurements, such as the jet transport coefficient $\hat{q}$~\cite{Burke:2013yra}. The interplay between theory and experiment on the single hadron production will help constraining the longitudinal distribution of parton energy loss in hot/dense QCD medium, and better understanding the jet-medium interactions after being combined with studies of full jets which also shed light on the angular distribution of the medium-induced gluon radiation and thus constrain the transverse distribution of parton energy loss as well~\cite{Vitev:2008rz,Vitev:2009rd,Dai:2012am,Aad:2010bu,Chatrchyan:2011sx,Kang:2014xsa}.

So far, most of the theoretical calculations on single hadron productions in HIC focus on $\pi$ meson or charged hadrons (where $\pi$ also giving a predominant contribution), and there are very few studies on other identified hadrons~\cite{Wang:1998bha,Liu:2006sf,Brodsky:2008qp,Chen:2008vha}. We note that $\eta$ meson is the second important source of decay electrons and photons just after the $\pi^0$, and its quantitative analysis is of importance to suppress the noise in measurements such as direct photon~\cite{Adler:2006hu,Abelev:2012cn}. In addition, PHENIX has made a detailed measurement of $\eta$ production in Au+Au collision at $\sqrt{s_{NN}}=200$~GeV, but there is not any theoretical calculations on $\eta$ at high $p_T$ in HIC to the best of our knowledge. It is of great interest to see how parton energy loss effect alters $\eta$ spectrum in HIC and whether the theoretical model of jet quenching could make a simultaneous description of both $\pi$ and $\eta$ production.

In this Letter we study $\eta$ productions at large $p_T$ in high-energy nuclear collisions at the RHIC and LHC in the framework of higher twist approach of jet quenching~\cite{guoxiaofeng,benwei-nuleon,Zhang:2003wk,Schafer:2007xh}. In higher twist approach, multiple scattering of the fast parton traversing through the dense QCD matter is calculated with generalized QCD factorization of twist-4 processes. The resulting parton energy loss due to multiple scattering leads to effectively modified parton fragmentation functions in medium (mFF). Incorporating mFFs into a next-to-leading order (NLO) perturbative QCD (pQCD) improved parton model, a phenomenological study on $\pi^0$ suppression in nucleus-nucleus collisions~\cite{Chen:2010te,Chen:2011vt} gives a decent description of $\pi^0$ yields in Au+Au collisions at the RHIC and in Pb+Pb reactions at the LHC. In this Letter, we may employ the identical model to investigate $p_T$ spectrum of $\eta$ production at NLO in HIC, with the same jet transport parameters extracted in $\pi^0$ production in HIC. We also explore the features of $\eta/\pi^{0}$ ratios in both $\rm p+p$ and $\rm A+A$ collisions. Although the flavor dependent parton energy loss in the QGP may alter the flavor compositions of fast partons which in turn have distinct probabilities fragmenting to $\eta$ and to $\pi$, at very high $p_{T}$, the predomination of quark contribution for both $\eta$ and $\pi^{0}$ ensures that the QCD medium effect does not affect the productions ratio of $\eta$ and $\pi^{0}$. The almost identical gluon (quark) contribution fractions to $\eta$ and to $\pi$ in $p+p$ may lead to a rather small deviation of $\eta/\pi^{0}$ in $\rm A+A$ with respect to that in $\rm p+p$ at intermediate and high $p_T$ region. And at small $p_T$, an enhancement of $\eta/\pi^{0}$ in $\rm A+A$ collisions can be seen because of the relative larger suppression of gluon contribution fraction to $\pi^{0}$  than that to $\eta$ in $\rm A+A$ collisions.

\hspace{0.7in}
\begin{figure}[!t]
\begin{center}
\hspace*{-0.1in}
\includegraphics[width=3.2in,height=2.8in,angle=0]{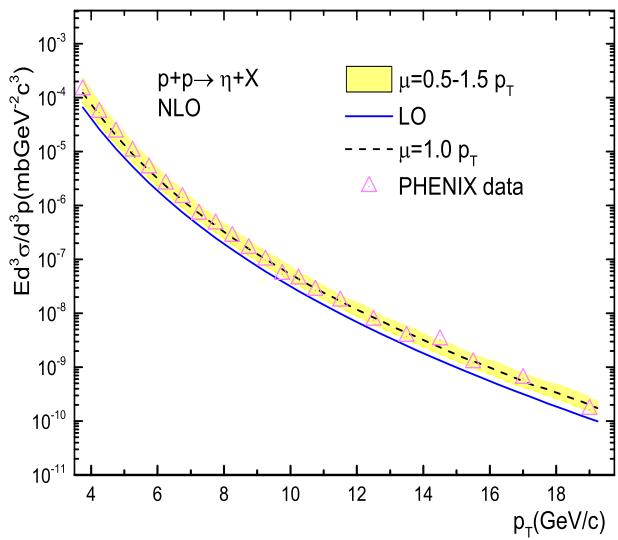}
\hspace*{-0.5in}
\caption{Comparison between the PHENIX data of $\eta$ invariance cross section in $\rm{p+p}$ collisions at
$200$~GeV and the NLO pQCD theoretical calculations.
}
\label{fig:illustpp}
\end{center}
\end{figure}
\hspace*{-1.5in}

Firstly, we investigate the single inclusive $\eta$ production in $p+p$ collisions, which provides the baseline for the nuclear modification of $\eta$ yield in HIC. A NLO pQCD improved parton model for the initial jet production spectra has been employed~\cite{Harris:2001sx}, in which the inclusive particle production cross section in $p+p$ collisions can be factorized into a convolution of parton distribution functions (PDFs) inside the proton, elementary partonic scattering cross sections $d\hat{\sigma}/d\hat{t}$, and parton fragmentation functions (FFs),
\begin{eqnarray}
\frac{d\sigma^h_{pp}}{dyd^2p_T}&=&\sum_{abcd}\int
dx_adx_bf_{a/p}(x_a,\mu^2)f_{b/p}(x_b,\mu^2) \nonumber \\
&&\hspace{-0.1in}\times\frac{d \hat {\sigma} }{d\hat{t}}(ab\rightarrow cd)
\frac{D_{h/c}^0(z_{c},\mu^2)}{\pi z_{c}}+\mathcal {O}(\alpha_s^3),
\label{eq:pp}
\end{eqnarray}
where $d \hat{\sigma}/d\hat{t}(ab\rightarrow cd)$ denotes leading order (LO) parton scattering cross sections at $\alpha_s^2$.
The NLO partonic cross section at at $\alpha_s^3$ includes $2\rightarrow3$ tree level contributions and one loop virtual corrections
to $2\rightarrow2$ tree processes~\cite{Owens}. In the above equation $f_{a/p}(x_a,\mu^2)$ stands for the PDFs in proton with $x_a$ the momentum fraction of the beam proton carried by the incoming parton, and in numerical simulations we use CTEQ6M parametrization~\cite{distribution}. $D^0_{h/c}(z_c,\mu^2)$ represents the parton fragmentation function in vacuum, which gives the possibility of parton $c$ fragmenting into hadron $h$ with momentum fraction $z_c$.  The factorization scale,
renormalization scale and fragmentation scale are usually chosen to be the same  and are related to
$p_T$ of the final hadron. This pQCD improved parton model at NLO accuracy has worked very nicely to describe data on inclusive pion production in p+p collisions at RHIC by using AKK pion FFs and the scale in the range $\mu=0.5 \sim 1.5 p_T$~\cite{Zhang:2007ja,Chen:2010te}.
\hspace{0.7in}
\begin{figure}[!t]
\begin{center}
\hspace*{-0.1in}
\includegraphics[width=3.4in,height=2.8in,angle=0]{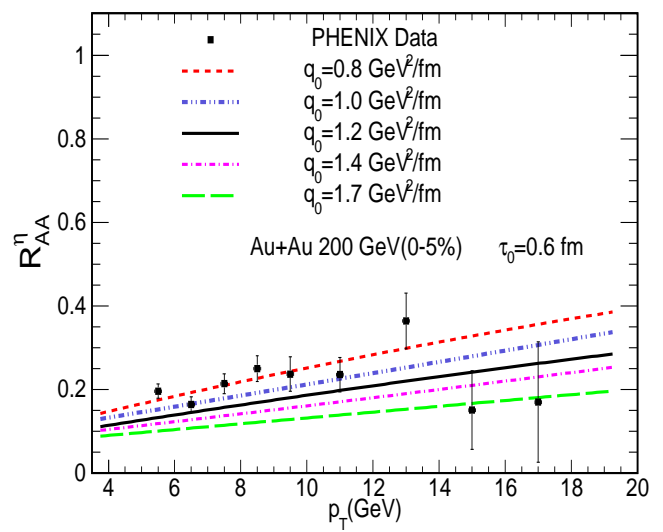}
\hspace*{-0.1in}
\caption{Comparison between the PHENIX Data of $\eta$ nuclear modification factor in $\rm{Au+Au}$ collisions at
$200$~GeV and numerical simulations at NLO.
}
\label{fig:illustetaraarhic}
\end{center}
\end{figure}
\hspace*{-0.5in}

To compute $\eta$ production  at LO and NLO with pQCD, the non-perturbative input of $\eta$ FFs will be needed.
The availability of a parametrization of $\eta$ FFs AESSS~\cite{Aidala:2010bn} allows us to make a NLO pQCD calculations for single-inclusive eta meson production as a function of final state $p_T$ in hadron-hadron collisions.
In AESSS parametrization because of the absence of enough data on inclusive $\eta$ productions, the $\eta$ FFs can not be extracted separately for each quark flavor without additional assumptions, and the assumption is made that all light quark fragmentations are the same. In Fig.~\ref{fig:illustpp} we confront our calculation with PHENIX data~\cite{Adare:2010cy}, and it is observed that the computed inclusive $\eta$ spectrum in $p+p$ collisions with the scale $\mu=1.0 p_T$ agrees well with the PHENIX data.

\hspace{0.7in}
\begin{figure}[!t]
\begin{center}
\hspace*{-0.1in}
\includegraphics[width=3.4in,height=2.8in,angle=0]{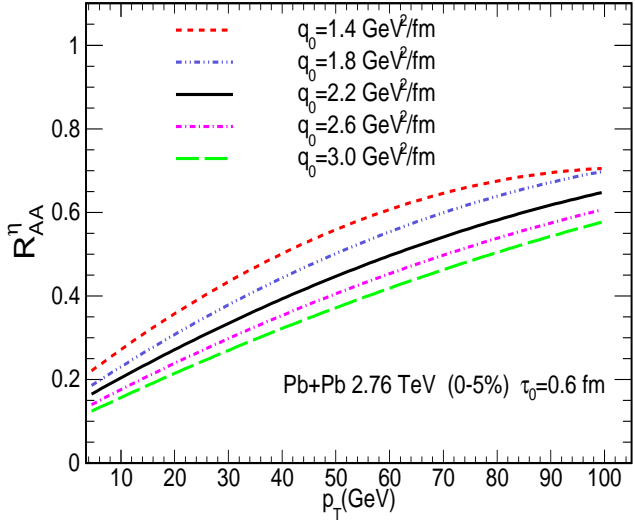}
\hspace*{-0.1in}
\caption{Predictions of $\eta$ medium modification factor in $\rm{Pb+Pb}$ collisions at
$2.76$~TeV with different $\hat{q}_0$ parameters.
}
\label{fig:illustetaraalhc}
\end{center}
\end{figure}
\hspace*{-1.5in}

In central high-energy nucleus-nucleus collisions, a hot and dense QCD matter may be created. A fast parton propagating in the QCD medium may encounter multiple scattering with other partons in medium and lose energy by induced gluon emission. In higher twist approach of
jet quenching~\cite{guoxiaofeng,benwei-nuleon,Zhang:2003wk,Schafer:2007xh}, this kind of multiple scattering is related to twist-4 processes of hard scattering and may give rise to the effective modifications of parton FFs in vacuum such as:
\begin{eqnarray}
\tilde{D}_{q}^{h}(z_h,Q^2) &=&
D_{q}^{h}(z_h,Q^2)+\frac{\alpha_s(Q^2)}{2\pi}
\int_0^{Q^2}\frac{d\ell_T^2}{\ell_T^2} \nonumber\\
&&\hspace{-0.7in}\times \int_{z_h}^{1}\frac{dz}{z} \left[ \Delta\gamma_{q\rightarrow qg}(z,x,x_L,\ell_T^2)D_{q}^h(\frac{z_h}{z},Q^2)\right.
\nonumber\\
&&\hspace{-0.2 in}+ \left. \Delta\gamma_{q\rightarrow
gq}(z,x,x_L,\ell_T^2)D_{g}^h(\frac{z_h}{z},Q^2) \right] ,
\label{eq:mo-fragment}
\end{eqnarray}
where $\Delta\gamma_{q\rightarrow qg}(z,x,x_L,\ell_T^2)$ and $\Delta\gamma_{q\rightarrow gq}(z,x,x_L,\ell_T^2)=\Delta\gamma_{q \rightarrow qg}(1-z,x,x_L,\ell_T^2)$ are the medium modified splitting functions calculated in higher twist approach~\cite{guoxiaofeng,benwei-nuleon}.
The medium-modified FFs averaged over the initial production position and jet propagation direction are given as~\cite{Chen:2010te,Chen:2011vt}:
\begin{eqnarray}
\langle \tilde D_{a}^{h}(z_h,Q^2,E,b)
\rangle &=& \frac{1}{\int
d^2{r}t_{A}(|\vec{r}|)t_{B}(|\vec{b}-\vec{r}|)} \nonumber \\
&&\hspace{-1.0in}\times \int\frac{d\phi}{2\pi}d^2{r}
t_{A}(|\vec{r}|)t_{B}(|\vec{b}-\vec{r}|)\tilde{D}_{a}^{h}(z_h,Q^2,E,r,\phi,b). \nonumber \\
\label{eq:frag}
\end{eqnarray}
where $b$ is the impact parameter and $t_{A,B}$ are the nuclear thickness functions in Glauber model~\cite{Miller:2007ri}. In the higher twist approach, one assumes that the parent energetic parton loses its energy in the QCD medium prior to vacuum hadronizations and single hadrons at large $p_T$ are fragmented in vacuum by the partons passed through the QCD medium.

Therefore, cross section of the single hadron in HIC collisions could be expressed as:
\begin{eqnarray}
\frac{1}{N_{\rm bin}^{AB}(b)}\frac{d\sigma_{AB}^h}{dyd^2p_T} &=&\sum_{abcd}\int
dx_adx_b f_{a/A}(x_a,\mu^2)f_{b/B}(x_b,\mu^2) \nonumber \\
&&\hspace{-0.5in}\times \frac{d\sigma}{d\hat{t}}(ab\rightarrow
cd)\frac{\langle \tilde{D}_{c}^{h}(z_{h},Q^2,E,b)\rangle}{\pi z_{c}}+\mathcal {O}(\alpha_s^3). \nonumber \\
\label{eq:AA}
\end{eqnarray}
Here $N_{\rm bin}^{AB}(b)$ gives the number of binary nucleon-nucleon collisions at the impact parameter $b$ in
$A+B$ collisions, $f_{a/A}(x_{a},\mu^{2})$ represents the effective PDFs inside a nucleus. In our calculations, we 
employed EPS09 NLO nuclear PDFs to include initial-state cold nuclear matter effects~\cite{Eskola:2009uj}.

\hspace{0.7in}
\begin{figure}[!t]
\begin{center}
\hspace*{-0.1in}
\includegraphics[width=3.4in,height=2.8in,angle=0]{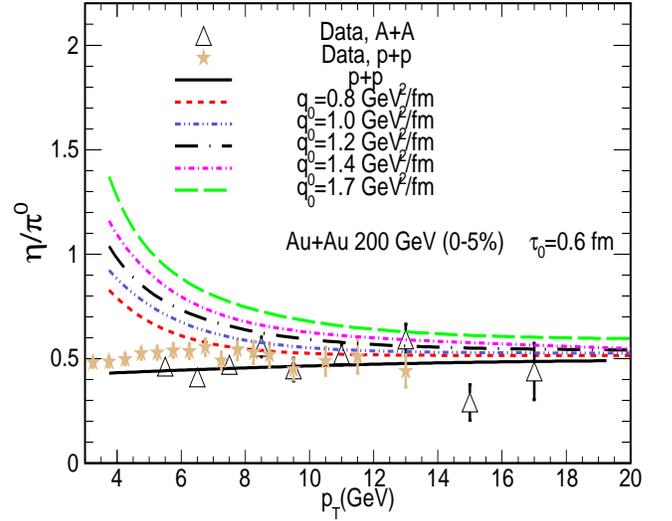}
\hspace*{-0.1in}
\caption{Comparison between the PHENIX data of $\eta/\pi^0$ ratio in$\rm{p+p}$ collisions and $\rm{Au+Au}$ collisions at
$200$~GeV and the numerical simulations at NLO.
}
\label{fig:illustetapirhic}
\end{center}
\end{figure}
\hspace*{-1.5in}

We assume all the energy loss of a fast parton is that carried away by the radiative gluon in the multiple scattering processes, the corresponding parton energy loss in the QCD medium can be expressed as:
\begin{eqnarray}
\frac{\Delta E}{E}=\frac{N_{c}\alpha_{s}}{\pi} \int dy^{-}dzd\ell_T^2\frac{(1+z)^3}{\ell_T^4}  \nonumber  \\
                            \times \hat{q}_{\rm R}(E, y) \sin^2[\frac{y^{-}\ell_T^2}{4Ex(1-z)}]
\label{eq:eloss}
\end{eqnarray}
which is proportional to the jet transport parameter $ \hat{q}_{\rm R}(E, y) $.  The jet transport parameter is related to the parton density distribution in the medium, therefore the space-time profile of the jet transport parameter characterizes the medium properties in the medium modified fragmentation functions and energy loss calculation.  Phenomenologically, we treat it proportional to the parton density relative to their values at the center of overlapped region in the most central collisions at initial time and a given temperature in an ideal hydro or in hadronic phase, the four momentum of the jet and the four flow velocity in the collision frame along the jet propagation path are also included~\cite{Chen:2010te}.

A full three-dimensional(3+1D)ideal hydrodynamics~\cite{Hirano2001,HT2002} to describe the space-time evolution of the QCD medium in heavy-ion collisions is employed to give the required space-time evolutionary informations of the medium such as: parton density, temperature, fraction of the hadronic phase, four flow velocity. In this theoretical framework, there left only one parameter when describe the medium modification of the hadron productions in heavy ion collisions:~$\hat{q}_0\tau_0$. In this Letter, we fix the initial time of the QGP medium at $\tau_0=0.6$~$fm$, and determine the initial values of the jet transport parameter $\hat{q}_0$ by fitting the final charged hadron multiplicity density in mid-rapidity of the hydrodynamics results~\cite{Chen:2010te,Chen:2011vt}. The $\hat{q}_0$ can also be considered to adjust the strength of jet-medium interaction as a parameter, the larger $\hat{q}_0$ is, the stronger the jet-medium interaction will be at every space-time points.

\hspace{0.7in}
\begin{figure}[!t]
\begin{center}
\hspace*{-0.1in}
\includegraphics[width=3.4in,height=2.8in,angle=0]{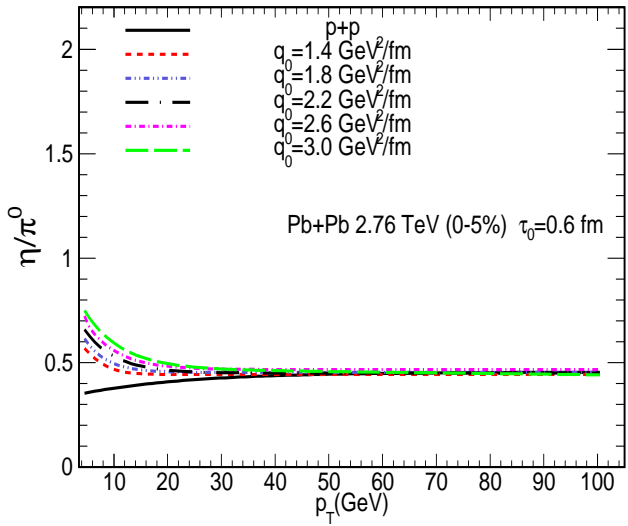}
\hspace*{-0.1in}
\caption{The NLO pQCD theoretical prediction of $\eta$ /$\pi^0$ ratio in $\rm{Pb+Pb}$ collisions at
$2.76$~TeV.
}

\label{fig:illustetapilhc}
\end{center}
\end{figure}
\hspace*{-1.5in}

With all these ingredients, we can calculate the hadron productions in heavy ion collisions up to the NLO.  Although we can directly compare the calculated hadron productions in heavy ion collisions with the experimental data, we introduce the nuclear modification factor $R_{AA}$ to better demonstrate and understand the medium modification phenomenon:
\begin{eqnarray}
R_{AB}(b)=\frac{d\sigma_{AB}^h/dyd^2p_T}{N_{bin}^{AB}(b)d\sigma_{pp}^h/dyd^2p_T}
\label{eq:eloss}
\end{eqnarray}

At various values of $\hat q_{0} \tau_{0}=0.48-1.02$ GeV$^2$,we calculate the $Au+Au$ productions within medium modified $\eta$ fragmentation functions in the $5\%$ most central $Au+Au$ collisions at RHIC energy $\sqrt{s}=200$~GeV, and compared our results with  the PHENIX experimental data on $R_{AA}$ of the $\eta$ spectra. The theoretical calculation results explained the data well in large $p_T$ region in Fig.~\ref{fig:illustetaraarhic}. The well agreement with the PHENIX data~\cite{Adler:2006hu} at mid-rapidity in the range $p_T = 2-20$~GeV shows that, even $\eta$ meson is 4 times heavier than $\pi^0$, a similar flat production suppression has been observed at RHIC in this $p_{T}$ range independent of their mass.

The prediction of $R_{AA}$ of the $\eta$ spectra in the central $Pb+Pb$ collisions at the LHC energy $\sqrt{s}$~=~$2.76$~TeV is given in  Fig.~\ref{fig:illustetaraalhc}. The $\hat q_{0}\tau_{0}$ are chosen from $0.84-1.8$ GeV$^{2}$. The suppression factor increases with $p_T$ was observed, it is mainly due to the energy dependence of parton energy loss and the less steep initial jet production spectra~\cite{Wang:2004yv}.

The $p_T$ dependence of the $\eta/\pi^0$ ratios are also investigated. We plot the $p_T$ dependence of the $\eta/\pi^0$ ratios in $Au+Au$ at $200$~GeV as in Fig.~\ref{fig:illustetapirhic}, and a good agreement between the model calculations with PHENIX data can be seen.  We also predict the $p_T$ dependence of the $\eta/\pi^0$ ratios in $Pb+Pb$ at $2.76$~TeV in Fig.~\ref{fig:illustetapilhc}.  Similar trend could be seen at the RHIC and LHC that with the increasing of $p_{T}$, the $\eta/\pi^0$ ratio in $A+A$ collisions comes closer to the $p+p$ curve, and at very larger $p_T$ two curves coincide with each other.

\begin{figure}[!t]
\begin{center}
\hspace*{-0.1in}
\includegraphics[width=3.4in,height=4in,angle=0]{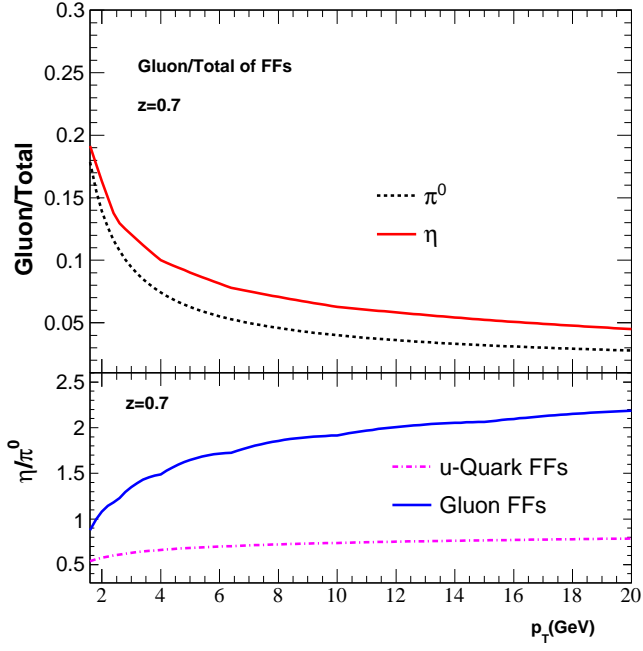}
\hspace*{-0.1in}
\caption{Top panel: gluon portion in total FFs (top panel) at fixed $z=0.7$; Bottom panel: $\eta/\pi^{0}$ ratio of the gluon, quark fragmentation functions (bottom panel) at fixed $z=0.7$.}

\label{fig:etapiffs}
\end{center}
\end{figure}

Due to the different color factors of quark-quark vertex and gluon-gluon vertex, in QCD medium gluon jet suffers larger energy loss than quark jet. Because quark FFs and gluon FF into $\eta$ and $\pi$ have quite different features (for instance, the bottom panel of Fig.~\ref{fig:etapiffs} shows that at very high $p_{T}$ region, $D_{q\to \eta}/D_{q\to \pi^{0}}$ at $z_h=0.7$ is  approximately $0.5$), in principle, a change of flavor compositions of parton jets may affect the ratio of $\eta/\pi^0$~\cite{Wang:1998bha}.  The coincidence of the $A+A$ and the $p+p$  $\eta/\pi^{0}$ curves in a wide $p_T$ region can not be explained in one simple story that parton jets loss their energies first in the QCD medium and then fragment into hadrons in the vacuum~\cite{Adler:2006hu}.

\hspace{0.7in}
\hspace*{-1.5in}

\begin{figure}[!t]
\begin{center}
\hspace*{-0.1in}
\includegraphics[width=3.4in,height=4in,angle=0]{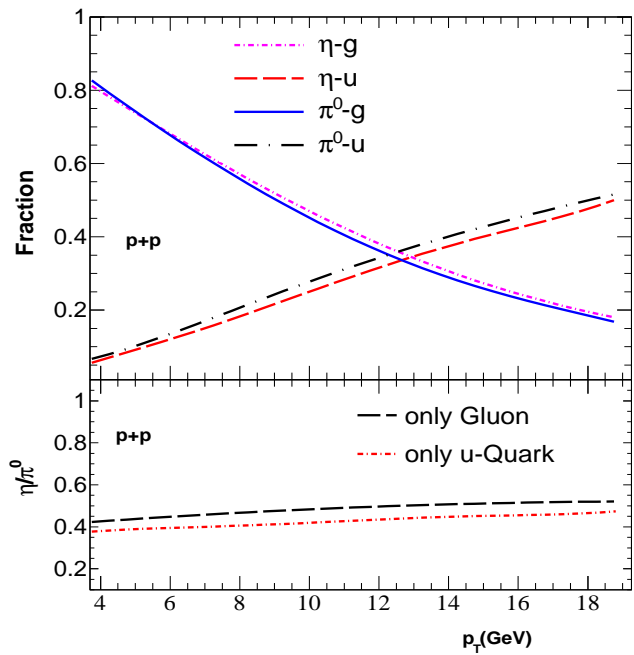}
\hspace*{-0.1in}
\caption{Top panel: quark and gluon contribution fractions to total $\pi^{0}$ (or $\eta$ meson) yields at NLO in $\rm p+p$ collisions at $200$~GeV.
Bottom panel: the ratio $\eta/\pi^{0}$ at NLO when only gluon or up quark contribution is considered in $\rm p+p$ collisions at $200$~GeV.}

\label{fig:etapifraction}
\end{center}
\end{figure}

To have a better understanding of the ratio of different hadron yields, we simplify the formula of identified hadron $p_{T}$ yield in $p+p$ collisions as:
\begin{eqnarray}
\frac{1}{p_{T}}\frac{d\sigma_{\pi^{0}, \eta}}{dp_{T}}=\int f_{q}(\frac{p_{T}}{z_{h}})\cdot D_{q\to \eta, \pi^{0}}(z_{h}, p_{T})\frac{dz_{h}}{z_{h}^2} \nonumber  \\
+ \int f_{g}(\frac{p_{T}}{z_{h}})\cdot D_{g\to \eta, \pi^{0}}(z_{h}, p_{T})\frac{dz_{h}}{z_{h}^2}  \,\,\, .
\label{eq:ptspec}
\end{eqnarray}
The above equation shows that the hadron yield in $\rm p+p$ will be determined by two factors: the initial (parton-)jet spectrum $f_{q,g}(p_T)$ and the parton fragmentation functions $D_{q,g\to \eta,\pi^{0}}(z_{h}, p_{T})$.  We plot the gluon's portion in total parton FFs to $\eta$ and to $\pi^{0}$ at fixed $z=0.7$ in Fig.~\ref{fig:etapiffs}(top panel).
we can see that, at very high $p_{T}$, (up-)quark FF is much larger than gluon FF for both $\eta$ and $\pi^{0}$ and shows weak $p_{T}$ dependence. At very high $p_T$, the yields of $\pi^0$ and $\eta$ will be almost only given by quark fragmentation (also see top panel of Fig.~\ref{fig:etapifraction}), and we can neglect the contributions of gluons. Noticing that, at high $p_T$, quark FFs $D_{q\to \eta, \pi^{0}}(z_h,Q=p_T)$ have a weak dependence on $z_{h}$ and $p_T$ in the typical $z_h$ region
$0.4-0.7$ for identified hadron production~\cite{Adler:2006sc}.
In the asymptotic region with $p_T \rightarrow \infty$ we obtain
\begin{eqnarray}
R({\eta/\pi^0})&=&\frac{d\sigma_{ \eta}}{dp_{T}}/\frac{d\sigma_{ \pi^0}}{dp_{T}} \nonumber \\
 &\approx&
\frac{\int f_{q}(\frac{p_{T}}{z_{h}})\cdot D_{q\to \eta}(z_{h}, p_{T})\frac{dz_{h}}{z_{h}^2}}
{\int f_{q}(\frac{p_{T}}{z_{h}})\cdot D_{q\to  \pi^{0}}(z_{h}, p_{T})\frac{dz_{h}}{z_{h}^2}}   \nonumber \\
&\approx&  \frac{\Sigma_{q} D_{q\to \eta}(\left<z_{h}\right>, p_{T})}
{\Sigma_q D_{q\to  \pi^{0}}(\left<z_{h}\right>, p_{T}) }   \,\,  .
\label{eq:ratio@high pT}
\end{eqnarray}
Therefore, even though quarks and gluons may lose different amount of energies and the flavor content of parton-jets will be changed in $A+A$ collisions, at the very high $p_T$ region in $\rm A+A$ collisions, the yields of both $\pi^0$ and $\eta$ should also predominantly come from quarks, and the ratio $\eta/\pi^0$ will also be determined only by quark FFs in vacuum with $z_h$ shift due to energy loss effect. As we mentioned before, in this kinematic region quark FFs have a weak dependence on $z_h$ and $p_T$, so we reach the conclusion that, at very high $p_T$ region, the ratios of $\eta/\pi^0$ in both $\rm A+A$ and $\rm p+p$ should overlap with the one in $e^+e^-$ scattering, and reach a universal value $\sim 0.5$.

Next, we consider the feature of $\eta/\pi^{0}$ when the transverse momentum $p_T$ is not very high. We define
\begin{eqnarray}
G^{\eta, \pi^{0}}(p_T)= \frac{\int f_{g}(\frac{p_{T}}{z_{h}})\cdot D_{g\to \eta, \pi^{0}}(z_{h}, p_{T})\frac{dz_{h}}{z_{h}^2} }
{ \frac{1}{p_{T}}\frac{d\sigma_{\pi^{0}, \eta}}{dp_{T}}} \,\,\, ,
\label{eq:ratio@gluon+quark}
\end{eqnarray}
which denotes the gluon contribution fraction to $\eta$ and $\pi^{0}$ yields with Eq.~(\ref{eq:ptspec}).
We plot the gluon (quark) contribution fraction to $\eta$ and $\pi^{0}$ yields in $p+p$ collision in Fig.~\ref{fig:etapifraction} (top panel). It shows that the quark contribution indeed dominate the $\eta$ and $\pi^{0}$ productions at larger $p_{T}$ region. Surprisingly, we also observe that
$$ G^{\pi^{0}}(p_T) \approx G^{\eta}(p_T) $$ in the $p_T$ region of $4-20$~GeV in $\rm p+p$ at RHIC.

From Eq.~(\ref{eq:ptspec}) we derive
\begin{eqnarray}
R({\eta/\pi^0})
&=&\frac{\frac{1}{1-G^{\eta}(p_T)}
\int f_{q}(\frac{p_{T}}{z_{h}})\cdot D_{q\to \eta}(z_{h}, p_{T})\frac{dz_{h}}{z_{h}^2}}
{\frac{1}{1-G^{\pi^0}(p_T)}
\int f_{q}(\frac{p_{T}}{z_{h}})\cdot D_{q\to  \pi^{0}}(z_{h}, p_{T})\frac{dz_{h}}{z_{h}^2}}   \nonumber \\
&\approx&
\frac{\int f_{q}(\frac{p_{T}}{z_{h}})\cdot D_{q\to \eta}(z_{h}, p_{T})\frac{dz_{h}}{z_{h}^2}}
{\int f_{q}(\frac{p_{T}}{z_{h}})\cdot D_{q\to  \pi^{0}}(z_{h}, p_{T})\frac{dz_{h}}{z_{h}^2}}   \nonumber \\
&\approx& \frac{\int f_{g}(\frac{p_{T}}{z_{h}})\cdot D_{g\to \eta}(z_{h}, p_{T})\frac{dz_{h}}{z_{h}^2}}
{\int f_{g}(\frac{p_{T}}{z_{h}})\cdot D_{g\to  \pi^{0}}(z_{h}, p_{T})\frac{dz_{h}}{z_{h}^2}}
 \,\,  .
\label{eq:ratio@general}
\end{eqnarray}
It implies that whereas $\pi^0$ and $\eta$ yields in $\rm p+p$ come from quark or gluon hadronization, when calculating the ratio $\eta/\pi^0$, because of the relative identical fractional contributions of gluon and quark to $\pi^0$ and $\eta$, we can consider the contributions of only quarks (or gluons). The flavor compositions or mixture of quarks and gluons in $\rm p+p$
have nearly negligible effect on $\eta/\pi^0$, which is confirmed by the numerical results shown in the bottom panel of Fig.~\ref{fig:etapifraction}.

\hspace{0.7in}
\hspace*{-1.5in}

\hspace{0.7in}
\begin{figure}[!t]
\begin{center}
\hspace*{-0.1in}
\includegraphics[width=3.4in,height=4in,angle=0]{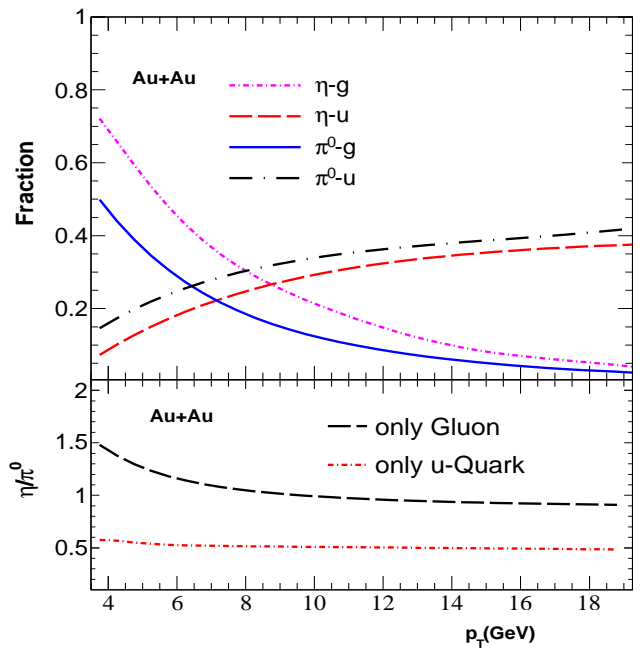}
\hspace*{-0.1in}
\caption{Top panel: quark and gluon contribution fractions to total $\pi^{0}$ (or $\eta$ meson ) yields  at NLO in $\rm Au+Au$ collisions at $200$~GeV.
Bottom panel: the ratio $\eta/\pi^{0}$ at NLO when only gluon or up quark contribution is considered in $\rm Au+Au$ collisions at $200$~GeV.}

\label{fig:etapifracAA}
\end{center}
\end{figure}
\hspace*{-1.5in}

Now we turn to $\eta/\pi^0$ in $\rm A+A$ collisions. Though flavor dependent energy loss in hot QCD medium may change the flavor compositions of jets, i.e., gluons may suffer more suppression than quarks, the effect to $\eta/\pi^0$ will be minimized by
the nearly identical $G^{\pi^{0}}(p_T)$ and $G^{\eta}(p_T)$ observed in $\rm p+p$, which underlies the overlapping of $\eta/\pi^{0}$ in $\rm A+A$ and the one in $\rm p+p$ in a wide region of $p_T$.

Nevertheless, small modifications to $\eta/\pi^0$ due to jet quenching effect will still exist.  In $\rm A+A$ the relation between $G^{\pi^{0}}(p_T)$ and $G^{\eta}(p_T)$ will be broken a little bit.
We plot the gluon and quark contribution fractions to $\eta$ and $\pi^{0}$ yields in $Au+Au$ collisions at $200$~GeV in Fig.~\ref{fig:etapifracAA}.  It reveals that the flavor dependent jet quenching effect really results in smaller gluon contribution fraction in heavy-ion collisions as compared to $\rm p+p$ in Fig.~\ref{fig:etapifraction}. A naive expectation is that because gluon may give larger $\eta/\pi^0$ ratio than quark does (see bottom panel of Fig.~\ref{fig:etapiffs}), the larger suppression of gluons in the QCD medium will reduce $\eta/\pi^0$. However, contrary to this naive expectation the jet quenching effect may instead enhance the ratio $\eta/\pi^0$ in high-energy nucleus-nucleus collisions as seen in
Fig.~\ref{fig:illustetapirhic} and Fig.~\ref{fig:illustetapilhc}.
The reason is that the suppression of gluon in QCD medium imposes a larger reduction of
the yield of $\pi^{0}$ than that of $\eta$ (top panel of Fig.~\ref{fig:etapifracAA}), thus giving rise to a slightly enhanced $\eta/\pi^0$ ratio in
$\rm A+A$. To demonstrate this we plot in the bottom panel of Fig.~\ref{fig:etapifracAA} the $\eta/\pi^0$ ratio in heavy-ion collisions originated from only gluon or quark contribution. One can see that a larger  $\eta/\pi^0$ ratio with only gluons included in the theoretical simulations. We emphasize that
 the identified hadron yield in heavy-ion collisions relies on three factors: the initial hard jet spectrum, the energy loss mechanism, and parton fragmentation functions to the hadron in vacuum. When $\eta/\pi^0$ ratios or other ratios of hadron yields are concerned , a simple estimation with only last two factors such as (flavor dependent) energy loss effect and FFs in vacuum may sometimes reach a wrong conclusion.

{\bf Acknowledgments:} This research is supported in part by  MOST in China under Project Nos. 2014CB845404, 2014DFG02050, and
by Natural Science Foundation of China with Project Nos. 11322546, 11435004, 11221504, 11347005 .

\end{document}